\documentclass[twocolumn,letter,appendixfloats]{aastex61}
\usepackage{graphicx}
\usepackage[fleqn]{amsmath}


\shorttitle{\sc{toros} observations of \sc{gw}170817}
\shortauthors{D\'{\i}az, Macri, Garcia Lambas, et al.}

\begin{document}
\title{Observations of the first electromagnetic counterpart to\\a gravitational wave source by the TOROS collaboration}

\author[0000-0002-7555-8856]{M.~C.~D\'{\i}az}
\affiliation{Center for Gravitational Wave Astronomy and Department of Physics \& Astronomy, University of Texas - R\'{\i}o Grande Valley, Brownsville, TX, USA}

\author[0000-0002-1775-4859]{L.~M.~Macri}
\affiliation{George P.~and Cynthia W.~Mitchell Institute for Fundamental Physics \& Astronomy, Department of Physics \& Astronomy, Texas~A\&M~University, College Station, TX, USA}

\author{D.~Garcia Lambas}
\affiliation{IATE-OAC, Universidad Nacional de C\'ordoba-CONICET, C\'ordoba, Argentina}

\author{C.~Mendes de Oliveira}
\affiliation{Instituto de Astronomia, Geof\'{\i}sica e Ci\^encias Atmosf\'ericas da U.~de S\~{a}o Paulo, S\~{a}o Paulo, SP, Brazil}

\author{J.~L.~Nilo Castell\'on}
\affiliation{Instituto de Investigaci\'on Multidisciplinario en Ciencia y Tecnolog\'{\i}a, Universidad de La Serena, La Serena, Chile}
\affiliation{Departamento de F\'{\i}sica y Astronom\'{\i}a, Universidad de La Serena, La Serena, Chile}

\author{T.~Ribeiro}
\affiliation{Departamento de F\'{\i}sica, Universidade Federal de Sergipe, S\~{a}o Crist\'ov\~{a}o, SE, Brazil}

\author{B.~S\'anchez}
\affiliation{IATE-OAC, Universidad Nacional de C\'ordoba-CONICET, C\'ordoba, Argentina}

\author{W.~Schoenell}
\affiliation{Instituto de Astronomia, Geof\'{\i}sica e Ci\^encias Atmosf\'ericas da U.~de S\~{a}o Paulo, S\~{a}o Paulo, SP, Brazil}
\affiliation{Departamento de F\'{\i}sica, Universidade Federal de Santa Catarina, Florian\'opolis, SC, Brazil}

\author{L.~R.~Abramo}
\affiliation{Departamento de F\'{\i}sica Matem\'atica, Instituto de F\'{\i}sica, Universidade de S\~{a}o Paulo, S\~{a}o Paulo, SP, Brazil}

\author{S.~Akras}
\affiliation{Departamento de Astronomia, Observat\'orio Nacional, Rio de Janeiro, RJ, Brazil}

\author{J.~S.~Alcaniz}
\affiliation{Departamento de Astronomia, Observat\'orio Nacional, Rio de Janeiro, RJ, Brazil}

\author{R.~Artola}
\affiliation{IATE-OAC, Universidad Nacional de C\'ordoba-CONICET, C\'ordoba, Argentina}

\author{M.~Beroiz}
\affiliation{Center for Gravitational Wave Astronomy and Department of Physics \& Astronomy, University of Texas - R\'{\i}o Grande Valley, Brownsville, TX, USA}

\author{S.~Bonoli}
\affiliation{Centro de Estudios de F\'{\i}sica del Cosmos de Arag\'on, Teruel, Spain}

\author{J.~Cabral}
\affiliation{IATE-OAC, Universidad Nacional de C\'ordoba-CONICET, C\'ordoba, Argentina}

\author{R.~Camuccio}
\affiliation{Center for Gravitational Wave Astronomy and Department of Physics \& Astronomy, University of Texas - R\'{\i}o Grande Valley, Brownsville, TX, USA}

\author{M.~Castillo}
\affiliation{Center for Gravitational Wave Astronomy and Department of Physics \& Astronomy, University of Texas - R\'{\i}o Grande Valley, Brownsville, TX, USA}

\author[0000-0002-2558-0967]{V.~Chavushyan}
\affiliation{Instituto Nacional de Astrof\'{\i}sica, \'Optica y Electr\'onica, Tonantzintla, Puebla, M\'exico}

\author{P.~Coelho}
\affiliation{Instituto de Astronomia, Geof\'{\i}sica e Ci\^encias Atmosf\'ericas da U.~de S\~{a}o Paulo, S\~{a}o Paulo, SP, Brazil}

\author{C.~Colazo}
\affiliation{IATE-OAC, Universidad Nacional de C\'ordoba-CONICET, C\'ordoba, Argentina}

\author{M.~V.~Costa-Duarte}
\affiliation{Instituto de Astronomia, Geof\'{\i}sica e Ci\^encias Atmosf\'ericas da U.~de S\~{a}o Paulo, S\~{a}o Paulo, SP, Brazil}

\author{H.~Cuevas Larenas}
\affiliation{Departamento de F\'{\i}sica y Astronom\'{\i}a, Universidad de La Serena, La Serena, Chile}

\author{D.~L.~DePoy}
\affiliation{George P.~and Cynthia W.~Mitchell Institute for Fundamental Physics \& Astronomy, Department of Physics \& Astronomy, Texas~A\&M~University, College Station, TX, USA}

\author{M.~Dom\'{\i}nguez Romero}
\affiliation{IATE-OAC, Universidad Nacional de C\'ordoba-CONICET, C\'ordoba, Argentina}

\author[0000-0001-5756-8842]{D.~Dultzin}
\affiliation{Instituto de Astronom\'{\i}a, Universidad Nacional Aut\'onoma de M\'exico, Ciudad de M\'exico, M\'exico}

\author{D.~Fern\'andez}
\affiliation{Instituto de Astrof\'{\i}sica, Pontificia Universidad Cat\'olica de Chile, Santiago, Chile}

\author{J.~Garc\'{\i}a}
\affiliation{Center for Gravitational Wave Astronomy and Department of Physics \& Astronomy, University of Texas - R\'{\i}o Grande Valley, Brownsville, TX, USA}

\author{C.~Girardini}
\affiliation{IATE-OAC, Universidad Nacional de C\'ordoba-CONICET, C\'ordoba, Argentina}

\author{D.~R.~Gon\c calves}
\affiliation{Observatorio do Valongo, Universidade Federal do Rio de Janeiro, R\'{\i}o de Janeiro, RJ, Brazil}

\author{T.~S.~Gon\c calves}
\affiliation{Observatorio do Valongo, Universidade Federal do Rio de Janeiro, R\'{\i}o de Janeiro, RJ, Brazil}

\author{S.~Gurovich}
\affiliation{IATE-OAC, Universidad Nacional de C\'ordoba-CONICET, C\'ordoba, Argentina}

\author{Y.~Jim\'enez-Teja}
\affiliation{Departamento de Astronomia, Observat\'orio Nacional, Rio de Janeiro, RJ, Brazil}

\author{A.~Kanaan}
\affiliation{Departamento de F\'{\i}sica, Universidade Federal de Santa Catarina, Florian\'opolis, SC, Brazil}

\author{M.~Lares}
\affiliation{IATE-OAC, Universidad Nacional de C\'ordoba-CONICET, C\'ordoba, Argentina}

\author{R.~Lopes de Oliveira}
\affiliation{Departamento de F\'{\i}sica, Universidade Federal de Sergipe, S\~{a}o Crist\'ov\~{a}o, SE, Brazil}
\affiliation{X-ray Astrophysics Laboratory and CRESST, NASA Goddard Space Flight Center, Greenbelt, MD, USA}

\author[0000-0002-1381-7437]{O.~L\'opez-Cruz}
\affiliation{Instituto Nacional de Astrof\'{\i}sica, \'Optica y Electr\'onica, Tonantzintla, Puebla, M\'exico}

\author{J.~L.~Marshall}
\affiliation{George P.~and Cynthia W.~Mitchell Institute for Fundamental Physics \& Astronomy, Department of Physics \& Astronomy, Texas~A\&M~University, College Station, TX, USA}

\author{R.~Melia}
\affiliation{IATE-OAC, Universidad Nacional de C\'ordoba-CONICET, C\'ordoba, Argentina}

\author{A.~Molino}
\affiliation{Instituto de Astronomia, Geof\'{\i}sica e Ci\^encias Atmosf\'ericas da U.~de S\~{a}o Paulo, S\~{a}o Paulo, SP, Brazil}

\author[0000-0001-9850-9419]{N.~Padilla}
\affiliation{Instituto de Astrof\'{\i}sica, Pontificia Universidad Cat\'olica de Chile, Santiago, Chile}

\author{T.~Pe\~ nuela}
\affiliation{Center for Gravitational Wave Astronomy and Department of Physics \& Astronomy, University of Texas - R\'{\i}o Grande Valley, Brownsville, TX, USA}
\affiliation{Ludwig Maximilian Universit\"at Munich, Faculty of Physics, Munich, Germany}

\author{V.~M.~Placco}
\affiliation{Department of Physics, University of Notre Dame, Notre Dame, IN, USA}
\affiliation{Joint Institute for Nuclear Astrophysics - Center for the Evolution of the Elements, USA}

\author{C.~Qui\~ nones}
\affiliation{IATE-OAC, Universidad Nacional de C\'ordoba-CONICET, C\'ordoba, Argentina}

\author{A.~Ram\'{\i}rez Rivera}
\affiliation{Departamento de F\'{\i}sica y Astronom\'{\i}a, Universidad de La Serena, La Serena, Chile}

\author{V.~Renzi}
\affiliation{IATE-OAC, Universidad Nacional de C\'ordoba-CONICET, C\'ordoba, Argentina}

\author{L.~Riguccini}
\affiliation{Observatorio do Valongo, Universidade Federal do Rio de Janeiro, R\'{\i}o de Janeiro, RJ, Brazil}

\author[0000-0002-4436-221X]{E.~R\'{\i}os-L\'opez}
\affiliation{Instituto Nacional de Astrof\'{\i}sica, \'Optica y Electr\'onica, Tonantzintla, Puebla, M\'exico}

\author{H.~Rodriguez}
\affiliation{IATE-OAC, Universidad Nacional de C\'ordoba-CONICET, C\'ordoba, Argentina}

\author{L.~Sampedro}
\affiliation{Instituto de Astronomia, Geof\'{\i}sica e Ci\^encias Atmosf\'ericas da U.~de S\~{a}o Paulo, S\~{a}o Paulo, SP, Brazil}

\author{M.~Schneiter}
\affiliation{IATE-OAC, Universidad Nacional de C\'ordoba-CONICET, C\'ordoba, Argentina}

\author{L.~Sodr\'e}
\affiliation{Instituto de Astronomia, Geof\'{\i}sica e Ci\^encias Atmosf\'ericas da U.~de S\~{a}o Paulo, S\~{a}o Paulo, SP, Brazil}

\author{M.~Starck}
\affiliation{IATE-OAC, Universidad Nacional de C\'ordoba-CONICET, C\'ordoba, Argentina}

\author{S.~Torres-Flores}
\affiliation{Departamento de F\'{\i}sica y Astronom\'{\i}a, Universidad de La Serena, La Serena, Chile}

\author{M.~Tornatore}
\affiliation{IATE-OAC, Universidad Nacional de C\'ordoba-CONICET, C\'ordoba, Argentina}

\author{A.~Zadro\. zny}
\affiliation{Center for Gravitational Wave Astronomy and Department of Physics \& Astronomy, University of Texas - R\'{\i}o Grande Valley, Brownsville, TX, USA}

\correspondingauthor{Lucas M.~Macri}
\email{lmacri@tamu.edu}

\begin{abstract} 
We present the results of prompt optical follow-up of the electromagnetic counterpart of the gravitational-wave event GW170817 by the Transient Optical Robotic Observatory of the South Collaboration (TOROS). We detected highly significant dimming in the light curves of the counterpart ($\Delta g=0.17\pm0.03$~mag, $\Delta r=0.14\pm0.02$~mag, $\Delta i=0.10\pm0.03$~mag) over the course of only 80 minutes of observations obtained $\sim 35$~hr after the trigger with the T80-South telescope. A second epoch of observations, obtained $\sim59$~hr after the event with the EABA 1.5m telescope, confirms the fast fading nature of the transient. The observed colors of the counterpart suggest that this event was a ``blue kilonova'' relatively free of lanthanides.
\end{abstract} 

\section{Introduction}

The network of advanced ground-based gravitational-wave (GW) interferometers constituted by the Advanced Laser Interferometer Gravitational-wave Observatory \citep[LIGO;][hereafter, ``LSC'']{LSC2015} started its second observational campaign (O2) on 2016 November 30. On 2017 August 1, Advanced Virgo \citep{Acernese2015} began its first observational campaign, initiating the first concurrent monitoring of the sky by a network of three GW interferometers\footnote{This concurrent campaign ended on 2017 August 25, \url{www.ligo.org/news/index.php\#O2end}}.

The first detection of a binary black hole (BBH) merger by Advanced LIGO opened the era of GW astronomy \citep[GW150914;][]{Abbott2016a}. Three similar events have been detected since then, two by Advanced LIGO and the most recent one by Advanced LIGO/Virgo \citep[GW151226, GW170104 and GW170814][]{Abbott2016b, Abbott2017a,Abbott2017b}. For many years, particularly since the discovery of the binary pulsar PSR B1913+16 by \citet{Hulse1975} and the evidence for energy loss in this system as expected from GW emission \citep{Taylor1982}, binary neutron star (BNS) mergers were anticipated to be one of the main sources found by advanced GW detectors. Hence, it was surprising that the first four GW detections were BBHs.

The LSC and the Virgo Collaboration (VC) issued on 2013 June 6 a worldwide call to participate in electromagnetic (EM) and multi-messenger observations of GW events recorded by their detectors, using a wide range of telescopes and instruments of ``mainstream astronomy''\footnote{\url{www.ligo.org/scientists/GWEMalerts.php}}. The Transient Optical Robotic Observatory of the South Collaboration \citep[TOROS;][]{Diaz2014,Benacquista2014}  was organized in 2013 to participate in these observations. While seeking to deploy a wide-field optical telescope on Cord\'on Mac\'on in north-west Argentina \citep{Renzi2009, Tremblin2012}, the collaboration has been utilizing other resources for follow-up activities. Our activities during LIGO's first observational campaign (O1) have been reported by \citet{Colazo2015} and \citet{Diaz2016}.

On 2017 August 17 12:41:04 UTC (BJD 2457983.02857), a BNS merger candidate was identified in data from the LIGO Hanford (H1) detector \citep[LIGO/Virgo G298048,][]{GCN21505,GCN21509}. The Gamma-ray Burst Monitor (GBM) onboard {\it Fermi} \citep{Bissaldi2009} detected an event $\sim2$~s after the GW trigger, which was given the designation GRB170817A \citep{GCN21506,GCN21520,GCN21528}. The significance of the GW detection was initially estimated as an equivalent false-alarm rate of 1-in-$10^4$~years based on the H1 data alone. The effective distance was estimated as $\sim58$~Mpc and the initial localization estimate, based only on H1 data, was quite broad. \citet{GCN21509} reported that the GW event was also clearly visible in data from LIGO Livingston (L1), although there was a coincident noise artifact. Further analysis of GW data from all detectors (including Advanced Virgo) provided better estimates of the localization probability and of the luminosity distance, $40\pm8$~Mpc \citep{GCN21510,GCN21513,GCN21527}.

At the time of the event, the horizon ranges \citep[the maximum distance at which a BNS merger could be detected with $S/N\!>\!8$;][]{Finn1993,Allen2012,Chen2017} were 218, 107, and 58~Mpc for L1, H1, and Virgo, respectively \citep{Abbott2017b}. The GW trigger has subsequently been confirmed as a very high-confidence detection consistent with a BNS merger, and given the designation GW170817 \citep{LVC2017}.

About 11~hr after the GW trigger, several groups participating in the aforementioned GW/EM collaboration reported \citep{GCN21529,GCN21530,GCN21531} the detection of a putative EM counterpart (initially called SSS17a or DLT17ck, official IAU name AT2017gfo; hereafter, ``the transient'') located $10\arcsec$ from the center of NGC$\,$4993, an S0 galaxy at a distance of $38\pm5$~Mpc \citep{Kourkchi2017}. Its J2000 coordinates were reported as R.A.=13h09m48.1s, Dec.=$-23$d22m53s \citep[][and D.~A.~Coulter et al.~2017, in prep.]{GCN21529}.

\begin{figure*}[t]
\includegraphics[width=\textwidth,angle=0]{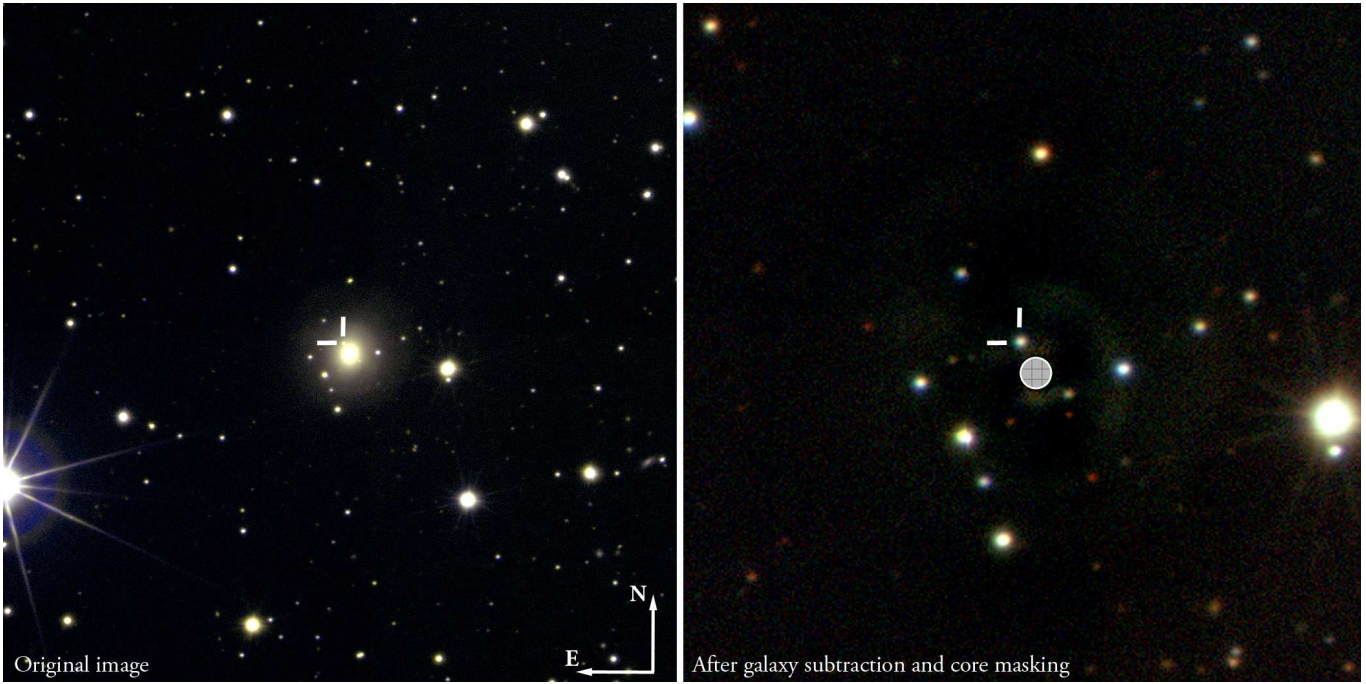}
\caption{Left: pseudo-color image of a small subsection ($9\farcm 5$ on a side) of the FoV of T80S, centered on the transient. Intensity scaling is logarithmic in order to better display the light distribution of the host galaxy. Right: $3\times$ zoom into the residual image after host galaxy subtraction and core masking (hatched circle; see \S\ref{sec:phot} for details). \label{fig:fov}}
\end{figure*}

\section{Observations and photometry} \label{sec:obs}

\subsection{Observations}

The TOROS Collaboration participated in the search for the EM counterpart of GW170817 starting just a few hours after the trigger. On the nights of 2017 August 17 and 18, we surveyed 26 nearby galaxies contained in the initial localization region using two facilities: the T80-South telescope (described below) and a Meade LX200 16-inch telescope equipped with a SBIG STF 8300 camera, located in Tolar Grande, Argentina \citep{GCN21619,GCN21620}. Once the candidate counterpart was identified near NGC$\,$4993 (see the references above), we focused our efforts on that source.

On 2017 August 18 ($\sim 35$~hr after the GW trigger) we observed the transient using the T80-South\footnote{\url{www.splus.iag.usp.br/t80s-telescope}} telescope (T80S; C.~Mendes de Oliveira et al.~2017, in prep.) located at CTIO. This telescope has a primary mirror with a diameter of 0.83m and a camera equipped with an E2V CCD290-99 detector, consisting of $9216\times9232$ pixels with a plate scale of $0\farcs 55/$pix that yields a field of view (FoV) $1\fdg 4$ on a side. We obtained 16, 15, and 15 one-minute exposures through SDSS $g$, $r$ and $i$ filters, respectively, at airmass values from 1.35 to 1.97 over the course of 80 minutes. The typical seeing was $\sim 1\farcs 8$. The left panel of Figure~\ref{fig:fov} shows a color composite of a small subsection ($9\farcm 5$ on a side) of the T80S FoV centered on the transient.

On 2017 August 19 ($\sim 59$~hr after the GW trigger) we imaged the same source using the 1.54m telescope located at the Estaci\'on Astrof\'{\i}sica de Bosque Alegre (EABA) and an Apogee ALTA F16 camera equipped with a KAF-16083 sensor, consisting of $4096\times4096$ pixels with a plate scale of $0\farcs 24/$pix that yields a FoV of $17\arcmin$ on a side. We obtained 88 one-minute unfiltered exposures in $2\times2$ binned mode to expedite readout and match the seeing ($\sim 3\farcs 5$).

\subsection{Photometry\label{sec:phot}}

Our photometry is based on the observations obtained with the T80S and the EABA 1.5m telescopes. We used the IRAF\footnote{IRAF is distributed by the National Optical Astronomy Observatory, which is operated by the Association of Universities for Research in Astronomy (AURA) under cooperative agreement with the National Science Foundation.} {\tt CCDPROC} package to debias and flat-field the raw frames. We carried out time-series point-spread function (PSF) photometry using {\tt DAOPHOT/ALLSTAR} \citep{Stetson1987}, ALLFRAME \citep{Stetson1994} and related programs, kindly provided by P.~Stetson. The steps we performed closely follow those outlined in \citet{Macri2006} and \citet{Macri2015}. We modeled the PSFs by fitting a Moffat function with $\beta=2.5$ to $25-50$ bright and isolated stars in each image. The fitting radii were 4 and 6~pix, and the PSFs were defined out to 8 and 10~pix for T80S and EABA, respectively. The local background level for each star was determined using annuli from $8-10$ and $10-15$~pix for T80S and EABA, respectively.

We first used {\tt DAOPHOT} to detect sources in each image with a significance of $4\sigma$ or greater, identify bright isolated stars, and determine the PSFs. We then used {\tt ALLSTAR} to obtain preliminary PSF photometry for all of the sources. Next, we used {\tt DAOMATCH} and {\tt DAOMASTER} to derive robust geometric transformations between all of the images obtained with a given telescope and filter. We generated four ``master frames'' (one for each telescope and filter) by median-combining a subset of images with low background values and good seeing. 

Given the close proximity of the transient to its host galaxy, we used the {\tt IMFIT} package \citep{Erwin2015} to model and subtract its light distribution from each image. We first determined the best-fit parameters on the master frames, masking of the all stars within $30\arcsec$ of the galaxy center as well as its innermost $5\arcsec$ and fitting the intervening region using S\'ersic profiles. Once the best-fit parameters (ellipticity, position angle, S\'ersic index, effective radius) for a given band were determined from the master image, they were held fixed in the fitting process for each individual frame. We only allowed the intensity scaling to remain a free parameter, since the location of the galaxy center was already well constrained by the initial PSF photometry. The right panel of Figure~\ref{fig:fov} shows the outcome of this procedure. Removing the host galaxy improved the determination of the local sky value and therefore reduced the uncertainty in the PSF magnitudes of stars in its vicinity by factors of $\sim1.5,2.2$ and 2.0 in {\it gri}, respectively.

\begin{figure}[t]
\includegraphics[height=0.495\textwidth,angle=270]{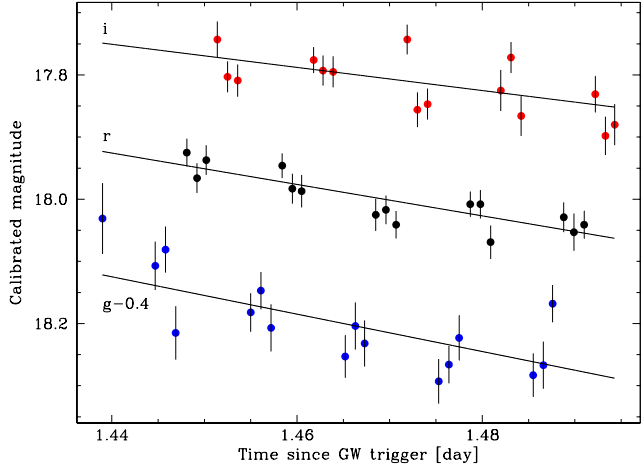}
\caption{{\it gri} light curves of the EM counterpart to GW170817, obtained with T80S on 2017 August 18. The $g$ points have been offset by $-0.4$~mag for clarity.\label{fig:lct80}}
\end{figure}

We performed the aforementioned procedures with {\tt DAOPHOT} and {\tt ALLSTAR} on the galaxy-subtracted master frames and generated ``master star lists'' for further analysis. The master star lists were used as input for {\tt ALLFRAME} to perform simultaneous PSF photometry on all of the host galaxy-subtracted images obtained with T80S in a given filter (due to the very low signal-to-noise ratio of the individual EABA images, we only obtained a single measurement for this data set using its master frame and did not perform time-series photometry). We identified 100 bright and isolated stars in each filter within the T80S FoV and used them to solve for frame-to-frame zeropoint offsets, in order to correct for differential extinction and any other variations. We achieved a photometric precision in our T80S time-series photometry of 0.01~mag or better for objects with $g<16,r<15,i<14$~mag. 

We transformed the T80S measurements into the Pan-STARRS1 photometric system \citep{Tonry2012} and simultaneously corrected for atmospheric extinction using $4600-5400$ objects in common between our star lists and the PS1 catalog available at the Mikulski Archive for Space Telescopes (MAST)\footnote{MAST is part of STScI, operated by AURA, Inc., under NASA contract NAS5-26555. Support for MAST for non-HST data is provided by the NASA Office of Space Science via grant NNX09AF08G and by other grants and contracts.}. We used the PS1 PSF magnitudes for the transformations. The selected stars spanned $-0.6 < g\!-\!i < 3.2$, enabling us to solve for a quadratic transformation in each band as a function of that color. We found small but statistically significant color terms for all of the transformations, with residual dispersions of $0.03$~mag.

The calibrated light curves are plotted in Figure~\ref{fig:lct80} and the time-series photometry is presented in Table~\ref{tab:lcdat} \footnote{Only a portion of the Table is shown here for guidance and context. The full version is available online.}. The tabulated uncertainties were calculated by {\tt ALLFRAME} and related programs based on the PSF fitting results and the frame-to-frame zeropoint corrections. We present the analysis of the light curves in \S\ref{sec:analysis}.

\begin{deluxetable}{lcrr}
\tablecaption{Time-series photometry\label{tab:lcdat}}
\tablewidth{0.495\textwidth}
%\tablehead{\colhead{Time$^a$\ \ \ \ } & \colhead{\ \ Band\ \ } & \colhead{\ \ Mag \ \ } & \colhead{\ \ $\sigma$ \ \ }}
\tablehead{\colhead{Time$^a$} & \colhead{Band} & \colhead{Mag} & \colhead{$\sigma$ (mag)}}
\startdata
1.4390 & g &\ \  18.43  0.06\\
1.4447 & g &\ \  18.51  0.04\\
1.4458 & g &\ \  18.48  0.04\\
1.4469 & g &\ \  18.62  0.04\\
1.4481 & r &\ \  17.93  0.02\\
1.4492 & r &\ \  17.97  0.02\\
1.4502 & r &\ \  17.94  0.02\\
1.4514 & i &\ \  17.74  0.03\\
%1.4525 & i &\ \  17.80 &\ \  0.03\\
%1.4536 & i &\ \  17.81 &\ \  0.03\\
\enddata
\tablecomments{$a$: days since GW trigger.}
\end{deluxetable}

We calibrated the EABA 1.5m observations in a similar manner. Due to the significantly smaller FoV and worse image quality, we were limited to 200 stars in common. Since these images were obtained without a filter, we solved for a linear transformation with respect to $r$, which exhibited a very large $r\!-\!i$ color term ($-0.54\pm0.03$).

\begin{figure}[t]
\includegraphics[height=0.495\textwidth,angle=270]{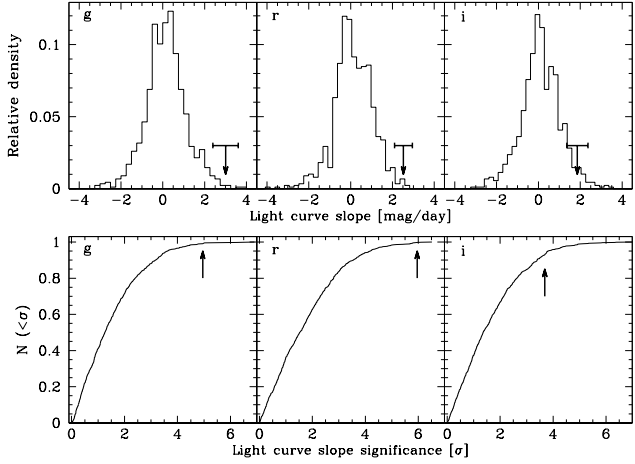}
\caption{Top: comparison of the light curve slopes (magnitudes per day) of the transient (arrows) in {\it gri} (left, center, and right panels, respectively) relative to 885 objects in the T80S FoV with similar colors ($|g-r|, |r\!-\!i|<0.1$). The horizontal error bars indicate the $1\sigma$ uncertainty in the values for the transient. Bottom: statistical significance of the values for the transient in {\it gri} (left, center, and right panels, respectively) relative to the comparison sample.\label{fig:decay}}
\end{figure}

\section{Analysis} \label{sec:analysis}

The T80S light curve of the transient exhibits a very significant decline across all bands during the $\sim 80$~minutes of observations. A weighted linear fit to the data yields $\Delta g=0.17\pm0.03$~mag, $\Delta r=0.14\pm0.02$~mag, $\Delta i=0.10\pm0.03$~mag over that time period. The mean magnitudes at the mid-point of our observations (1.467~day after the GW trigger) and their time derivatives (expressed in magnitudes per day) are as follows:
\begin{eqnarray*}
g = 18.60\pm0.02~{\rm mag,} & \hspace{0.2in} & dg/dt=3.0\pm0.6~{\rm mag/day}\\
r = 17.99\pm0.02,\hspace{0.31in} &  & dr/dt=2.5\pm0.4\\
i = 17.80\pm0.02,\hspace{0.33in} &  & di/dt=1.9\pm0.5.
\end{eqnarray*}

We ruled out systematic effects as the reason for the fast decline by examining the light curves of all 885 objects in the T80S FoV with similar colors to the transient (within 0.1~mag in both $g-r$ and $r\!-\!i$). We performed weighted linear fits as a function of time on the light curves of all selected objects in each band, and estimated the statistical significance of the first-order coefficient (hereafter, ``slope''). The results are shown as histograms in the top half of Figure~\ref{fig:decay}. It can be seen that the distributions are approximately centered on zero and that the transient exhibits some of the fastest recorded decline rates in all bands. The bottom panels of Fig.~\ref{fig:decay} show that most of the light curve slopes for the objects in the comparison sample are not statistically significant, while the decline rates for the transient are among the most highly statistically significant.

\begin{figure}[t]
\includegraphics[height=0.495\textwidth,angle=270]{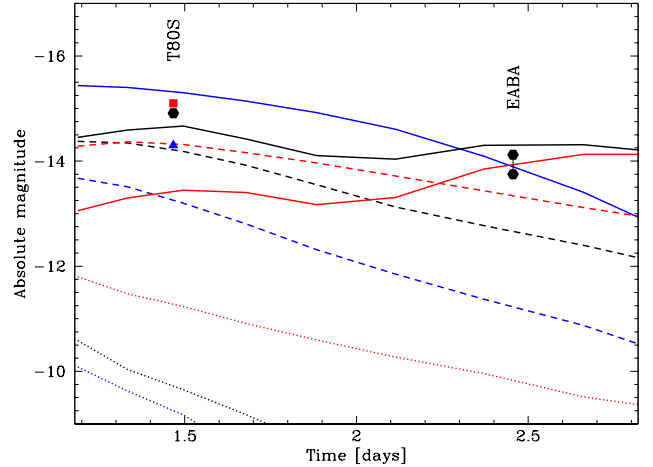}
\caption{Comparison of our photometry ($g$: blue triangle; $r$: black hexagons; $i$: red square) adjusted to $D=38$~Mpc with models from \citet{Tanaka2017} plotted using the same color scheme. The dotted lines represent a ``red kilonova'' model with dynamical ejecta rich in lanthanides. The dashed and solid lines represent ``blue kilonova'' wind models with decreasing amounts of lanthanides. The measurement uncertainties are smaller than the size of the symbols. The two possible $r$ values at 2.456~days are discussed in \S3.\label{fig:mod}}
\end{figure}

As we do not have a measurement of the $r\!-\!i$ color of the transient during the EABA observations, we present two possible values for the calibrated magnitude of the object at that time (2.456~days after the trigger): $r=18.78\pm0.03$~mag if its color did not evolve relative to the previous night ($r\!-\!i=0.19\pm0.03$~mag), or $r=19.15\pm0.06$~mag if its color evolved as extrapolated from the T80S light curve to $r\!-\!i\sim0.84\pm0.08$~mag. Despite the limited information obtained from EABA, these observations confirm a fast decline in luminosity of $\Delta r \sim0.8-1.2$~mag over 24~hr.

Figure~\ref{fig:mod} compares our photometry (adjusted to $D\!=\!38$~Mpc) with the {\it gri} light curves predicted by three kilonova models \citep[see Figure 8 of][hereafter ``T17'']{Tanaka2017}, which build upon the work of \citet{Tanaka2013}. We refer the interested readers to those publications for details of the parameters used in each model.

Our absolute magnitudes and colors are inconsistent with the predictions of a ``red kilonova'' model containing dynamical ejecta ($0.01 M_\odot$, $v=0.2c$) rich in lanthanides (dotted lines). Our $r$-band luminosity is in fairly good agreement with the prediction of a ``blue kilonova'' model with a ``wind'' ($0.01 M_\odot$, $v=0.05c$) free of lanthanides ($Y_e=0.3$ variant in T17, solid lines), but the predicted $g\!-\!i$ color of this model at $t=1.5$~days does not match the observations. A similar ``wind model'' ($0.01 M_\odot$, $v=0.05c$), in which the ejecta contain a small amount of lanthanides ($Y_e=0.25$ variant in T17, dashed lines), matches the observed colors fairly well, but the predicted luminosities are somewhat lower than observed. Regardless, it appears that this event may have been a ``blue kilonova'' \citep{Metzger2017} relatively free of lanthanides.

Intriguingly, there is a low-significance ($\sim 0.1$~mag) modulation in the T80S light curve residuals across all bands. If confirmed by others, such a correlated departure from a linear decay may be useful for discriminating among models. As seen in Fig.~\ref{fig:mod}, the light curves predicted by the two ``blue'' kilonova models exhibit quite different behavior at the time of our observations.

\section{Concluding Remarks} \label{sec:conclusion}

The detection of a merger of two neutron stars by Advanced LIGO/Virgo has opened a new chapter of GW astronomy. The combination of three GW interferometers provided a robust and relatively small localization region for the event, and the prompt follow-up and sharing of information by many collaborations resulted in a timely identification of its EM counterpart.

Given the fast temporal evolution of these transients, it is clear that continued worldwide follow-up efforts will be critical for providing early and nearly continuous coverage, minimizing weather losses, and maximizing the astrophysical constraints that can be extracted from photometric and spectroscopic observations.

\vspace{12pt}

\acknowledgments
We thank the referee for a very prompt and helpful report and Dr.~Masaomi Tanaka for kindly sharing his latest kilonova models in advance of publication. MCD acknowledges NSF support through grant NSF-HRD 1242090. DGL and the IATE/UNC team acknowledge support from the Consejo Nacional de Investigaciones Cient\'{\i}ficas y T\'ecnicas of Argentina. CMdO and the S-PLUS team are thankful to FAPESP (grant 2009/54202-8) for funding of the T80-South robotic telescope and its camera, to the Observatorio Nacional-MCT for funding the T80-South building, to INPE for help with the design of the camera, and to the staff of CEFCA for constant support over the years in the design and testing of T80S and its data reduction pipeline. JLNC is grateful for financial support from the Southern Office of Aerospace Research and Development (SOARD), a branch of the Air Force Office of the Scientific Research International Office of the United States (AFOSR/IO), through grant FA9550-15-1-0167. JLNC also acknowledges financial support from the Direcci\'on de Investigaci\'on y Desarrollo de la Universidad de La Serena through the Programa de Incentivo a la Investigaci\'on Acad\'emica (PIA-DIULS). RLO was supported by the Brazilian agency CNPq (PDE-200289/2017-9 and Universal-459553/2014-3). STF acknowledges financial support by the Universidad de La Serena for funding two technician positions for the T80S project. Finally, the entire TOROS Collaboration would like to thank the Argentine Gemini TAC and the Mexican GTC TAC for awarding ToO time to TOROS during the first semester of 2017.

\emergencystretch=3em
\bibliographystyle{aasjournal}

\end{document}